\documentclass[journal]{IEEEtran}
\ifCLASSINFOpdf
\else
\fi
\usepackage{graphicx}

\begin{document}
\title{Transmission Delay of Multi-hop Heterogeneous Networks for Medical Applications}

\author{M. M. Yaqoob, I. Israr, N. Javaid, M. A. Khan, U. Qasim$^{\ddag}$, Z. A. Khan$^{\S}$\\

        $^{\ddag}$University of Alberta, Alberta, Canada\\
        Department of Electrical Engineering, COMSATS\\ Institute of
        Information Technology, Islamabad, Pakistan. \\
        $^{\S}$Faculty of Engineering, Dalhousie University, Halifax, Canada.
        }

\maketitle

\begin{abstract}
Nowadays, with increase in ageing population, Health care market keeps growing. There is a need for monitoring of Health issues. Body Area Network consists of wireless sensors attached on or inside human body for monitoring vital Health related problems e.g, Electro Cardiogram (ECG), ElectroEncephalogram (EEG), ElectronyStagmography(ENG) etc. Data is recorded by sensors and is sent towards Health care center. Due to life threatening  situations, timely sending of data is essential. For data to reach Health care center, there must be a proper way of sending data through reliable connection and with minimum delay. In this paper transmission delay of different paths, through which data is sent from sensor to Health care center over heterogeneous multi-hop wireless channel is analyzed. Data of medical related diseases is sent through three different paths. In all three paths, data from sensors first reaches ZigBee, which is the common link in all three paths. After ZigBee there are three available networks, through which data is sent. Wireless Local Area Network (WLAN), Worldwide Interoperability for Microwave Access (WiMAX), Universal Mobile Telecommunication System (UMTS) are connected with ZigBee. Each network (WLAN, WiMAX, UMTS) is setup according to environmental conditions, suitability of device and availability of structure for that device. Data from these networks is sent to IP-Cloud, which is further connected to Health care center. Main aim of this paper is to calculate delay of each link in each path over multi-hop wireless channel.
\end{abstract}
\section{Introduction}
Many countries face ageing population, as number of senior citizens increasing all over the world. With increase in ageing population, there is a need to monitor their Health on regular basis. Specialized Health monitoring of serious cases are very important, however, it is quiet expensive. With emerging technology, remote patient monitoring is possible. WBAN consists of interconnected sensors, either placed around body or small enough to be placed inside the body. It provides ease of connectivity with other systems and networks thus allowing proper Health monitoring. With help of WBAN, monitoring of patient is done remotely through internet, intranet or any other network. These sensors continuously monitor data and send it to Health care center.\\
\indent\indent In patient monitoring system data transmission reliability with low delay is very important. Different technologies have been used in sending of medical data to Health care center like bluetooth connected to cellular system[1]. In this paper, we present three different paths, which can be employed at different places for the monitoring of serious critical data of ECG, EEG etc. Different devices connected with WBAN takes data from sensor nodes and transmits it to Health care center.\\
\indent A low cost and reliable approach is preferred for transmission of data from sensor nodes to Health monitoring room. Heterogeneous multi-hop transmission represents sending of medical data to Health monitoring room using multiple devices that are interconnected with each other, each performing its own operation. In this paper, we implement three different paths for sending of data. Each path is implemented according to need and environmental conditions. We calculate delay of all links that takes part in data sending for all three paths. ZigBee network is connected to nodes attached to body, for better monitoring and control.\\
\indent Bluetooth considered to be a network that has easy connection with cellular systems. Both ZigBee and Bluetooth offer low power consumption, respective data rate and both are useful for short range. However, potential interference from different devices is a concern for Bluetooth[1]. ZigBee consumes less power than Bluetooth. WLAN is connected with ZigBee, which is referred as path 1 for data transmission. It consists of 802.15 having easy connectivity and supports Wireless Fidelity (Wi-Fi). WiMAX is connected with ZigBee in path 2, WiMAX provides high data rate with long coverage area. Path 3 comprises of ZigBee connected with UMTS.\\
\indent In this work, we focus on calculating delays of all three paths, by inspecting delays of each link of each path. Firstly sensor nodes that are connected to body collects data of ECG, EEG or other medical applications and sends data to local relay device like ZigBee. Data then travels to second device connected with ZigBee either WLAN, WiMAX or UMTS, depending on need, suitability, or conditions. After that data is send to IP-Cloud, which is connected to Health care center.\\
\indent We consider three hop structure for each path, which consists of a ZigBee link, from sensor nodes to ZigBee device, second hop from ZigBee device to WLAN, WiMAX or UMTS and third hop from WLAN, WiMAX or UMTS to IP-cloud as shown in fig 1.\\
\indent The rest of the paper is organized as follows. Related Work and Motivation are given in section II. Section III describes Network Architecture for Heterogeneous Multi-Hop Networks in WBAN. Section IV concludes the work with future work.
\section{Related Work and Motivation}
Authors evaluate overall transmission delay of ECG packets over two-hop wireless channel.  ECG data from  BAN are compressed and sent through a Bluetooth-enabled ECG monitor to a smart phone and thereafter to a cellular Base Station (BS). Exploiting the inherent heartbeat pattern in ECG traffic, they introduce a context aware packetization for ECG transmission in [1]. However, ECG is not only data that can be transmitted, other medical data through different networks is sent. In this paper we calculate delay of heterogeneous networks of multi-hop which is not done in paper [1].
\\
\indent Authors in paper [2], state that, IEEE 802.15.4 standard is designed as a low power and low data rate protocol with high reliability. They analyze unslotted version of protocol with maximum throughput and minimum delay. The main purpose of IEEE 802.15.4 standard is to provide low power, low cost and highly reliable protocol. The standard defines a physical layer and a MAC sub layer. This standard operates in either beacon enabled or non beacon mode. Physical layer specifies three different frequency ranges, 2.4 GHz band with 16 channels, 915 MHz with 10 channels and 868 MHz with 1 channel. Calculations are done by considering only beacon-enabled mode and with only one sender and receiver. However, it has high cost of power consumption. As number of sender increases, efficiency of 802.15.4 decreases. Throughput of 802.15.4 declines and delay increases when multiple radios are used because of increase in number of collisions.
\\
\indent In paper [3], authors adopt a tree protocol for ECG signal carried over ZigBee. A prototype of DSP platform enabling good performance in ZigBee is established. Symmetrical system architecture is developed. They develop mathematical model and simulated transmission time delay of ECG data. Mathematical model is built for CSMA/CA mechanism. However authors consider ZigBee. As ZigBee is for low distance coverage and low data rate, we consider a heterogeneous network environment for sending of medical data through WLAN, WiMAX and UMTS networks.
\\
\indent WBAN is used to develop a patient monitoring system which offers flexibility and mobility to patients. It allows flexibility of using remote monitoring system via either internet or intranet. Performance of IEEE 802.15.4/ZigBee operating in different patient monitoring environment is examined in [4]. However, authors simulate hospital network based on Ethernet standard. Also authors not discussed mathematical aspects of their environment setup, which we do in our paper.
\\
\indent An application of wireless cellular technologies CDMA2000 1xEVDO, as a promising solution to wireless medical system is propose in [5]. Authors analyze end-to-end delay analysis for medical application using CDMA2000 1xEVDO. However, they only consider worst-case end to end delay over cellular network. They have not discussed about interoperability of CDMA2000 1xEVDO with WBAN. Also they only analyze mathematical equations for ECG data. In our paper, we analyze for general medical traffic.
\\
\indent Authors in [6], examine use of IEEE 802.15.4 standard in ECG monitoring sensor network and study the effects of CSMA/CA mechanism. They analyze performance of networks in terms of transmission delay, end to end latency, and packet delivery rate. For time critical applications, a payload size between 40 and 60 bytes is selected due to lower end to end latency and acceptable packet delivery rate. However, authors consider only single hop communication. We calculate and implement it with heterogeneous networks multi-hop transmission.
\\
\indent Authors present a new cross-layer communication protocol for WBANs,  CICADA in [7]. This protocol setup a network tree in a distributed manner. This structure is used to guarantee collision free access to the medium and to route data towards the sink. Energy consumption is low because nodes can sleep in slots, where they are not transmitting or receiving. However, their propose protocol only support node to sink traffic.
\\
\indent A Traffic-adaptive MAC protocol (TaMAC) is introduced by using traffic information of sensor nodes in [8]. TaMAC protocol is supported by a wakeup radio which is used to support emergency and on-demand events in a reliable manner. Authors compare TaMAC with beacon-enabled IEEE 802.15.4 MAC, WiseMAC, and SMAC protocols. They study co-existence of heterogeneous WBAN traffic. However, authors have not simulated delay in a heterogeneous environment.
\\
\indent Authors in [9] propose a MAC protocol for WBAN using wakeup radio mechanism. TDMA based scheme combined with wakeup radio is used to design an energy efficient Medium Access Control (MAC). However, their simulations are only for single hop communication. Moreover authors have not done calculations for heterogeneous network environment, as we do it in our paper.
\\
\indent In paper [10], authors discuss three emerging wireless standards, ZigBee, WiMAX and Wi-Fi. They are discussing in this paper, characteristics and application of these emerging technologies. Comparisons are presented to prove which performs better. However, no scenario is considered, in which all these technologies are implemented or combined with one another and nothing is proved graphically.
\section{Network Architecture for Heterogeneous Multi-hop Network in WBAN}
One of most important metrics for QoS is delay. There is a need to know delay of data passing through different paths and devices. It is highly dependent on type of structure used, protocols used in networks and requires high level of reliability. In this paper we calculate delay of all links that takes part in sending data from different device to the Health center.
\begin{eqnarray}
D_{total}= D1+D2+D3
\end{eqnarray}
$D_{total}= Total$ $delay$ $from$ $sensor$ $node$ $to$ $Health$ $monitoring$ $room$
\\
$D1=Delay$ $of$ $link 1$
\\
$D2=Delay$ $of$ $link 2$
\\
$D3=Delay$ $of$ $link 3$
\\
\begin{figure}[!h]
\centering
\caption{Heterogeneous Multi-hop Network in WBAN}
\includegraphics[width=3.5 in, height=2.5 in]{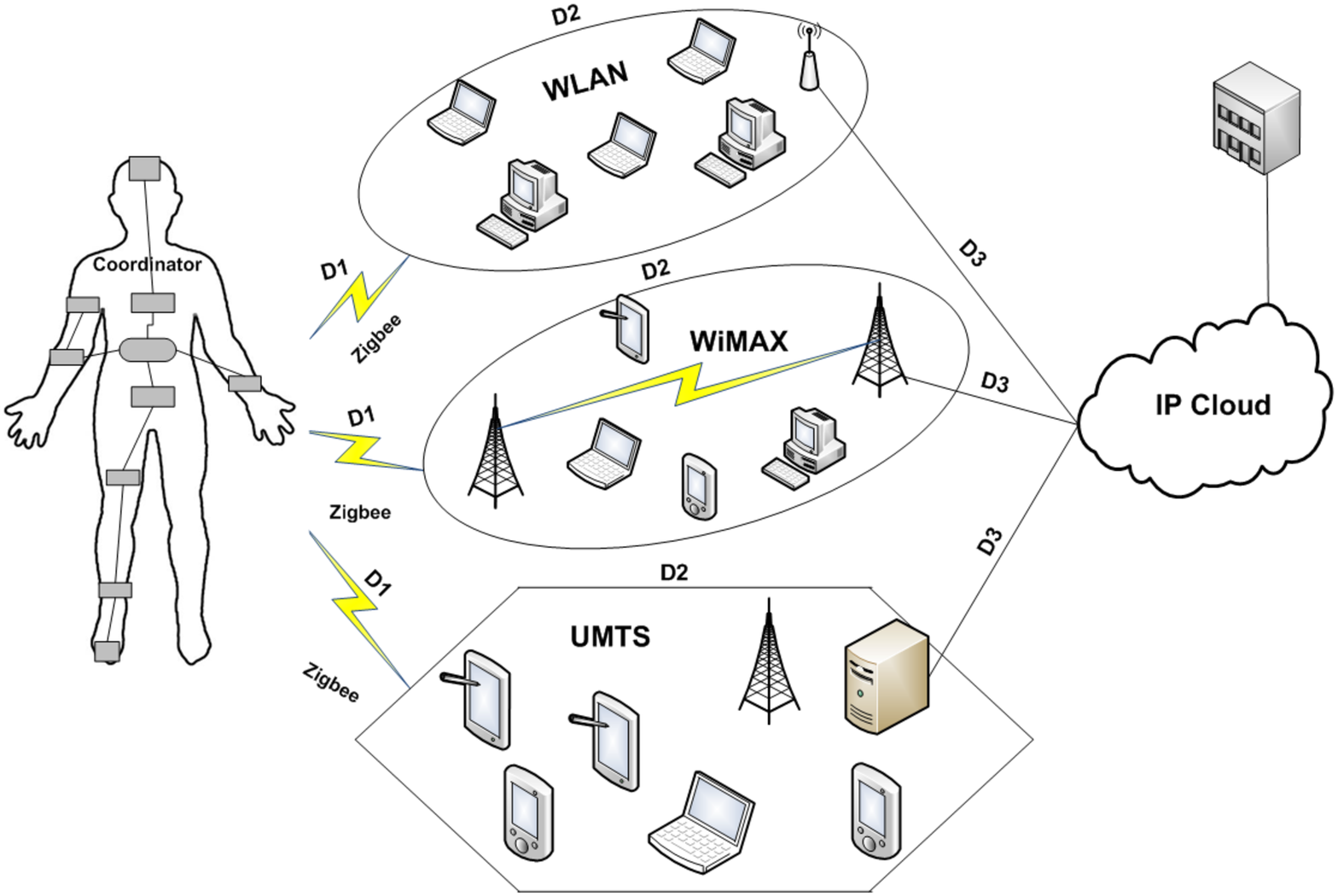}
\end{figure}
\\
Fig 1 shows complete structure of data transmission from three different paths to Health monitoring room. Total delay is calculated by equation 1. Delays of each link are shown in fig 1. Delay 1 (D1) is delay from nodes connected with human body to ZigBee, Delay 2 (D2) is from ZigBee to ext connected network WLAN, WiMAX or UMTS and Delay 3 (D3) is from WLAN, WiMAX or UMTS to IP-cloud and then from IP-cloud to Health monitoring room. \\
\indent In first path data travels from nodes attached to human body  towards ZigBee as shown in fig 1. After passing on to ZigBee, data divides according to path. If first path is established then data goes towards WLAN and devices attached to WLAN. If second path is implemented then data passes on to WiMAX. If path 3 is established then data passes on to UMTS.\\
\indent Delay of all these paths are referred as delay 2 according to the implementation shown in fig 1. After data is passed from any of the three networks WLAN, WiMAX, UMTS, it is send to IP-cloud which is connected to Health care center.\\
\indent All technologies are implemented according to need, ZigBee is used as starting device, because it is best short range network. After ZigBee, if we want to develop a Wi-Fi structure for sending of data to the monitoring room then, best device is WLAN which supports Wi-Fi internet with reliability and gives good coverage area. If we want to develop a last mile network and also have well developed structure of WiMAX than it is the network that is needed to be used. It supports high data rate and also gives last mile connectivity. If want to develop a cellular structure based on mobile traffic then after ZigBee, UMTS structure is established, which is based on cellular structure and supports all kind of mobile traffic with good coverage area. Handoff is a problem in UMTS, but it is minimized by using different techniques.
\subsection{ZigBee}
It is small, low-power digital radio based on IEEE 802.15.4 standard for Personal Area Networks. It is less expensive than other Wireless Personal Area Networks (WPANs), such as Bluetooth. ZigBee is used in those application where we need low data rate, long battery life and secure networking. It supports data rate up to 250 Kbit/s. It is best suited for periodic data or single signal transmission from sensor to other device input. Low cost of ZigBee allows it to be widely develop in wireless control and monitoring applications like sending of medical data from sensor node on human body to other devices. Its network layer supports both star, tree and mesh networks. ZigBee builds up MAC layer and Physical layer for low data rate WPANs.
\\
\indent Nodes in ZigBee go from sleep to active mode in 30ms or less due to which its latency is low. Due to long sleep time of nodes, its power consumption is low and gives long battery life. Its protocols are intended for low data rates and have low power consumption thus resulting network uses small amount of power. ZigBee devices are of three types:
\\
\indent $ZigBee$ $Coordinator$ (ZC):This is root of all network and makes bridges with other networks. There is exactly one ZigBee coordinator in each network because it starts the network.
\\
\indent $ZigBee$ $Router$ (ZR):It runs an application function and can act as router passing data from other devices.
\\
\indent $ZigBee$ $End$ $Device$ (ZED): It only communicates with the coordinator or router. Nodes in ZED are in sleep mode thereby giving long battery life. ZED requires least amount of memory and therefor it is less expensive than ZR and ZC.

\subsection{WLAN}
It links more than one device by using some wireless method. It connects through an access point to internet. This gives mobility to move around and still connected to internet. WLAN is based on IEEE 802.11 like ZigBee. It is most popular end user device due to cheap cost, ease of installation and wireless access to users.
\\
\indent Many projects are setup using WLAN to provide wireless access to different locations. WLAN supports data rate up to 54Mbit/s. It is long term and cost effective. It is easier to add more stations with WLAN. It provides connectivity in those areas where there are no cables.
\subsection{WiMAX}
It is a standard made for wireless communication providing 120 Mbit/s data. It easily passes range of WLAN, offering area network with signal radius of 50km. It provides data transfer rates that are superior to cable-modem and Digital Subscriber Line (DSL) connections. It is based on IEEE 802.16 family of networks standards. WiMAX is similar to Wi-Fi but provides much greater data rate and greater distance.
\\
\indent WiMAX internet connectivity is provided to multiple devices and it is connected to multiple devices to provide internet to home, business and other places. Standard 802.16 is versatile enough to accommodate Time Division Duplexing(TDD) and Frequency Division Duplexing(FDD).
\subsection{UMTS}
It is based on GSM and relates to 3rd generation of wireless technology. It uses Wideband Code Division Multiple Access (W-CDMA) to provide greater efficiency and bandwidth to mobile network operators. UMTS supports maximum data transfer rates up to 42Mbit/s.
\subsection{Heterogeneous Multi-hop Paths}
In all three parts, sensor node collects medical data from body and sends it to ZigBee, ZigBee being the first device connected to human body sensors in all three paths.
In path 1 sensor nodes, ZigBee, WLAN, IP-cloud and Health monitoring room are connected. Path 2 consists of sensor nodes, ZigBee, WiMAX, IP-cloud and Health monitoring room. Path 3 consists of sensor nodes, ZigBee, UMTS, IP-cloud and Health monitoring room. Sensor nodes are attached to body of a person, from where they record the medical data and sends it to ZigBee. ZigBee through end device, receives data and passes it to coordinator, so that, coordinator passes data to router from where data is sent to WLAN in path 1 and to WiMAX in path 2. ZigBee router, coordinator, and end device used in this modeling contains, application layer, network layer and IEEE 802.15.4 MAC standard. End device of ZigBee is configured to operate as router so it sends data to WLAN and WiMAX for further processing. Parameters are set according to the requirement. Packet Inter arrival Time is 0.04 seconds with packet size of 1024 bytes.
\\
\indent As data is passed from ZigBee to WLAN and WiMAX, both of them takes data through end device and passes it to its router from where it is send to server and then from there to IP-cloud. Data is also send to end device of WLAN and WiMAX. After sending of data to IP-cloud, data further processed to the Health monitoring room.
\\
\indent Structure of WLAN consists of Application layer, Data link layer, Transport layer and Physical layer. Data from all these layers is processed and then passed to IP-cloud. Data when reaches WiMAX, it passes through WiMAX user station, IP backbone and WiMAX base station(BS). As WiMAX is 802.16 family member so it consists of Application Layer, Transport Layer, Routing Layer and Data link layer.
\\
\indent IP-cloud is made by using custom model in OPNET and its parameters are set. Each device in this path is configured according to the need. Some of the paraments are kept constant for these devices.
\\
\indent As ZigBee operates on 802.15.4 hence it has CSMA/CA mechanism. Minimum backoff exponent is kept to 2, with maximum number of backoff to 3. Channel sensing duration is 0.1 seconds. Network parameters that are used in ZigBee coordinator and end device are given in table I.
\begin{table}
\caption {Network Parameters Of ZigBee coordinator and End device}
\begin {center}
\begin {tabular} {| p{2.5cm} | p{1cm} |}
\hline
Parameters type & Value \\ \hline
Beacon Order &   6 \\ \hline
Superframe Order & 0 \\ \hline
Maximum Routers & 5 \\ \hline
Maximum Depth & 5 \\ \hline
Beacon Enabled Network & Disabled \\ \hline
Mesh Routing  & Disabled \\ \hline
Route Discovery Timeout  & 10(msec) \\ \hline
Back off Exponent & 3 \\ \hline
Maximum Number Of Backoffs & 2\\ \hline
\end{tabular}
\end{center}
\end{table}
\\
\indent In this table beacon order of IEEE 802.15.4 is set to 6. When we look at structure of IEEE 802.15.4 we know about the beacon order[2]. Superframe order has also been explained in paper [2]. Maximum number of routers that can be connected to coordinator and end device is kept to 5 with each having depth of 5. Beacond enabled network is disabled for better performance. As there are not many devices that needs to be routed hence Mesh routing is also disabled.

\indent In case of WLAN there are many paraments that needed to be set before start of communication between WLAN and ZigBee, routing protocols in it are kept to default so that routing takes place as normally. IP is kept to constant, so that WLAN can easily communicate with IP-cloud. There are two WLAN MAC addresses that need to be set for this communication one is $IF0_{P0}$ and other one is $IF1_{PO}$. Both these addresses are of different LAN parameters that needed to be set. Parameters of these MAC addresses are given in table II.
\\
\begin{table}
\caption {WLAN PARAMETERS OF $IF0_{P0}$ and $IF1_{P0}$}
\begin {center}
\begin {tabular} {| p{2.5cm} | p{2cm} |}
\hline
Name of Parameter & Value \\ \hline
BSS Identifier & Auto Assigned \\ \hline
Physical technique &   Direct Sequence \\ \hline
Data Rate & 11Mbps \\ \hline
Transmit Power & 0.005 \\ \hline
Packet Reception threshold power & -95 dBm \\ \hline
Buffer size & 25600 bits \\ \hline
Large Packet Processing & Drop \\ \hline
\end{tabular}
\end{center}
\end{table}
\indent In this table Basic Service Set Identifier(BSSID) is the MAC address of station in an access point. It is administrated MAC address generated from 42 bit address. It is either kept to auto assigned so that WLAN assigns different BSSID to different stations or it is assigned by user. Each BSSID of device is separated from other, so that, conflict of same BSSID does not occur. Physical technique that is used in CSMA/CA is DSSS, input data is encrypted with a unique signal and then at the output it is decrypted using same code and data is recovered [2]. Data rate has been kept to 11Mbps for this device. Minimum power at which a node accepts a packet is referred as packet reception threshold power. Sending of large packet are set to drop, if there is a large packet waiting to be send, then it is dropped because large packet cause more delay and in this critical system we need packet with less delay.\\
\indent Parameter values in WiMAX are adjusted according to requirement. MAC address is auto assigned and IP routing protocols are kept to default. WiMAX contains base station(BS) parameters which include Code Division Multiple Access(CDMA), are kept to 8. Rest of the Parameters with their values are given in table III.
\begin{table}
\caption {WiMAX Parameters Value}
\begin {center}
\begin {tabular} {| p{2.5cm} | p{2cm} |}
\hline
Name of Parameter & Value \\ \hline
Antenna Gain & 15(dbi) \\ \hline
Number of Transmitters & SISO \\ \hline
MAC Address &   Auto Assigned \\ \hline
Maximum Transmission Power  & 0.5(W) \\ \hline
Physical profile & Wireless OFDM 20MHZ \\ \hline
Maximum number of SS nodes & 100 \\ \hline
Minimum Power Density & -110(dBm) \\ \hline
Maximum Power Density & -60(dBm) \\ \hline
\end{tabular}
\end{center}
\end{table}
\\
\indent In table III antenna gain of WiMAX is set to 15 dbi. MAC addresses are kept to auto assigned WiMAX auto assigns the MAC address of devices to avoid conflict of addresses. Transmission power that is be transmitted by antenna is 0.5W. Physical Profile which is used by WiMAX is OFDM at 20MHZ of frequency. \\
\indent In path three sensor nodes send packets containing information about Health to ZigBee and from ZigBee it is send to UMTS from where it is transferred to IP-cloud and Health monitoring room. ZigBee performs similar to the two path explained above and settings of parameters are also same. UMTS is a universal mobile telecommunication system which works on cellular technology. ZigBee is allowed to send data to UMTS by changing its inner structure, UMTS receives data from the end device of ZigBee, working as a router through Node B, than data is passed through Authentication Authorization and Accounting {AAA} server where authentication take place. Authorization confirms subscribers configuration information and finally collection billing information in done in Accounting.  After all this process data is passed to UMTS Node B.
\\
\indent UMTS structure contains Home Location Register (HLR). HLR is centralized database unit that contains all data of mobile phone connected to core network. It also stores data of every Subscriber Identification Number (SIM) card issued by companies to customers. It contains UMTS Serving GPRS Support Node (SGSN) which works as Mobile Switching Center, UMTS Radio Network Controller (RNC) which works as Base Station Controller (BSC) and UMTS Base Station (BS). Data from networks passes to the UMTS server connected to router which is connected to IP-Cloud after passing through BS, BSC, RNC, SGSN and HLR.
\\
\indent Server of UMTS contains application, transport, routing and  data link layer in its structure each performs its own operation and these layers are set according to use. UMTS HLR, RNC and Router consists of complex inner structure. UMTS parameter values are given in table IV.
\\
\begin{table}
\caption {UMTS Parameters value}
\begin {center}
\begin {tabular} {| p{2.5cm} | p{2cm} |}
\hline
Name of Parameter & Value \\ \hline
Processing time & 0.002 (sec) \\ \hline
Maximum retry on time expiry & 4 \\ \hline
Cell path loss parameters & Default \\ \hline
UMTS cell id & Default \\ \hline
UMTS SGSN ID & 0 \\ \hline
IP & Default \\ \hline
\end{tabular}
\end{center}
\end{table}
\indent Link one in all three paths and link two in path 1 works on CSMA/CA mechanism, so transmission delay in it is calculated by the equation 2. T is the total transmission time that is needed to send data from sender to receiver. Backoff time may vary up to 5 depending on the parameters of the network device and traffic load. Inter Frame Space is delay that comes after any type of communication in CSMA/CA. Turnaround time is the transmission time of sending a packet and then receiving acknowledgement successfully[2].
\begin{eqnarray}
D=T_{bo}+T_{data}+T_{ta}+T_{ack}+T_{ifs}
\end{eqnarray}

Data transmission time $T_{data}$, Backoff slots time $T_{bo}$, Acknowledgement time $T_{ack}$ are given by equation 2, 3, 4 respectively[2].
\\
\begin{eqnarray}
T_{data}=\frac{L_{phy} + L_{mac hdr} + payload + L_{mac ftr}}{R_{data}}\\
\nonumber\\
T_{bo}=bo_{slots} * T_{boslots}      \\
\nonumber\\
T_{ack}= \frac{L_{phy} + L_{mac hdr} + L_{mac ftr}}{R_{data}}
\end{eqnarray}
\\
Where
\\
T$_{bo}$=Back Off Period
\\
T$_{data}$=Transmission Time of Data
\\
T$_{ta}$=Turn Around Time
\\
T$_{ack}$=Acknowledgement Transmission Time
\\
T$_{ifs}$=Inter Frame Space
\\
T$_{phy}$=Length of Physical header
\\
L$_{mac hdr}$=Number of MAC header
\\
Payload=Number of data byte in packet
\\
L$_{mac ftr}$=size of MAC footer
\\
$bo_{slots}$=Number of back off slots
\\
$T_{bo slots}$= Time for a back off slot
\\

\indent In CSMA/CA mechanism packet losses due to collision. Collision occurs, when more than one node, sensing medium and finding it idle at same time. If acknowledgement time is not taken in to account then there is no retransmission of packet and it is considered that each packet is delivered successfully.
The probability of end device successfully transmitting a packet is modeled as follows[2].
\\
\begin{eqnarray}
P_{backoff period}=\frac{1}{2^{BE}}
\\
\nonumber\\
P_{tss}= \frac{1}{D}{(1-\frac{1}{D})}^{BE-2} \\
= p({1-p})^{BE-2} \nonumber
\end{eqnarray}
\indent Where D is the number of end devices that is connected to the router or coordinator. BE is the backoff exponent in our case it is 3.$P_{tss}$ is the probability of transmission success at a slot. $\frac{1}{D}$ is the probability of end device successfully allocated a wireless channel.\\
\indent General formula for $P_{time delay event}$ is given by equation 8. Probability of time delay caused by CSMA/CA backoff exponent is estimated as in [3]. Value of BE=3 is used in following estimation. Minimum value of BE is 2 and we estimate it by applying summation from 2 to 3. $P_{tde}$ is the probability of time delay event.
\\
\begin{eqnarray}
P_{tde}= \sum_{n=0}^{2^{BE-1}} n\frac{1}{2_{BE}}p{1-p}^{BE-2}
\end{eqnarray}
\begin{eqnarray}
P_{tde}= \sum_{n=0}^{3} n\frac{1}{2_{BE}}p + \sum _{n=4}^{11} n\frac{1}{2_{BE}}p
\end{eqnarray}
\indent Expectation of the time delay is obtained as follows[2]. $P{E_{A}}$ and $P{E_{B}}$ are taken from equations 8 and 9 respectively.
\\
\begin{eqnarray}
E[Time Delay]=P{E_{A}|E_{B}}\\
\nonumber\\
= \frac{\sum_{n=0}^{3} n\frac{1}{2_{BE}}p + \sum _{n=4}^{11} n\frac{1}{2_{BE}}p}{\sum_{n=0}^{2^{BE-1}} n\frac{1}{2_{BE}}p{1-p}^{BE-2}}
\end{eqnarray}
\\
\section{Conclusion and Future Work}
In this paper, delay performance of transmission of medical data from sensors connected to body or implanted, over a heterogeneous multi-hop environment through three different paths have been analyzed. Wireless transmission of medical data goes through different devices in each path, devices have been setup on the basis of environmental conditions, structure of network and availability of device setup. ZigBee is common network in each path which is connected in link 1, it receives data from sensor nodes and passes it to network connected in link 2. Link 2 of each path has a different device. Path 1 has WLAN, path 2 has WiMAX and path 3 has UMTS connected with ZigBee. After receiving data from ZigBee, these all three devices in link 2 sends data to IP-cloud connected in link 3. Data from IP-cloud is send to Health care center for monitoring of vital Health related issues. Delay of each link is calculated for each path. Based on our analysis, overall transmission delay of each path can be evaluated over three hop wireless channel. This work focuses on calculation of delay of each link connected in each path. Future work includes calculation of overall transmission delay, sending of data from nodes and receiving instruction from Health monitoring room through different paths.

\bibliographystyle{plain}

\end{document}